\documentclass[showpacs,amsmath,amssymb]{revtex4-2}
\usepackage{float}
\usepackage{tikz}
\usetikzlibrary{calc}
\usetikzlibrary{backgrounds}
\usepgflibrary{shapes}
\usetikzlibrary{through}
\usepackage{graphicx}
\usepackage{dcolumn}
\usepackage{bm}
\usepackage{xcolor}

\newcommand{\MDB}[1]{{#1}}

\begin{document}

\preprint{AIP/123-QED}

\title[Three- and four-wave resonances in the nonlinear quadratic Kelvin lattice]
{ Three- and four-wave resonances in the \MDB{nonlinear quadratic Kelvin lattice}}

\author{A. Pezzi}
\email{andrea.pezzi@unito.it}
\affiliation{
Dipartimento di Fisica, Universit\`a degli Studi di Torino, 10125 Torino, Italy
}%
\author{T. Comito}%
 \email{tiziana.comito@ucdconnect.ie}

\author{M. D. Bustamante}%
\email{Corresponding author: miguel.bustamante@ucd.ie }
 \homepage{http://maths.ucd.ie/~miguel}
\affiliation{ 
	School of Mathematics and Statistics, University College Dublin, Belfield, Dublin 4
}%

\author{M. Onorato}
\email{miguel.onorato@unito.it}
\affiliation{%
Dipartimento di Fisica, Universit\`a degli Studi di Torino, 10125 Torino, Italy and INFN, Sezione di Torino, 10125 Torino, Italy
}%

\date{\today}

\begin{abstract}
In this paper we investigate analytically and numerically the nonlinear Kelvin lattice, namely a chain of masses and nonlinear springs, \MDB{as in the $\alpha$-Fermi-Pasta-Ulam-Tsingou (FPUT) chain}, where, \MDB{in addition}, each mass is connected to a nonlinear resonator, {\it i.e.}, a second mass free to oscillate.  Both nonlinearities are quadratic in the equations of motion. This setup represents the simplest prototype of  nonlinear wave propagation on a nonlinear metamaterial. In the linear case, we diagonalize the system,  and the two branches of the dispersion relation  can be found.
Using this result, we derive in the nonlinear case  the equations of motion for the normal variables in Fourier space, obtaining a system governed by triad interactions among the two branches of the dispersion relation. We find that the transfer of energy between these two branches is ruled by three- and four-wave resonant interactions.  We perform numerical simulations of the primitive equations of motion and highlight the role of resonances as an efficient mechanism for transferring energy. Moreover, as predicted by the theory, we provide direct evidence that  four-wave resonances appear on a time scale that is longer than the time scale for three-wave resonances. \MDB{We also assess the recurrence behaviour (usual in the FPUT system) for the nonlinear Kelvin lattice, and we show that, while recurrence is observed if all the energy is placed, at time $t=0$, in the lowest mode of the acoustical branch, a non-recurrent behaviour is observed if the initial energy is located in the optical branch.}

\end{abstract}

\maketitle

\section*{INTRODUCTION}

Waves are ubiquitous in physics and engineering. Advances in technology have created an increasing demand for controlling waves that traditional materials cannot satisfy. 
In this context, metamaterials offer the possibility to manipulate some of the properties of waves such as their refraction or their dispersion, and are now the subject of many interdisciplinary studies \cite{zhu2018metamaterials}, \cite{liu2015metamaterials}. Metamaterials are artificially engineered structures that interact uniquely with waves; their properties depend on the geometrical construction, rather than on the chemical composition of the material. Metamaterials have several practical applications, among which are the absorption of mechanical vibrations \cite{al2022advances}, the manipulation of acoustic waves \cite{wang2022models}, and the prevention of damages from earthquakes  \cite{miniaci2021hierarchical} and coastal erosion \cite{de2021attenuating}. For classification and further developments, consult \cite{singh2015review,hussein2014dynamics}.

\MDB{In its simplest setting, a metamaterial can be modelled as a one-dimensional chain (masses and springs) to which one or more masses, acting as resonators, are included. The aim is to understand some basic physical properties of the models, such as the branching out of the dispersion relation, the formation of  frequency band gaps (i.e. ~some frequency ranges are forbidden, or alternatively they have negative effective mass), the nonlinear interactions between modes belonging to the different branches, and new low-order resonances that may arise. } Although the governing nonlinear evolution equations can be solved numerically with little effort, the chaos present in the system makes an analytical study necessary for a full comprehension of the properties of the metamaterial and for their exploitation. To this end, in the present paper we consider nonlinearities both in the propagating medium and in the resonator. We give a fully nonlinear treatment to the problem, emphasising the possibility of transferring energies among the normal modes of the dispersive curves. This approach is exploited for the simplest system considered (a dispersion relation with two branches), and it can be extended in a straightforward manner (although algebra becomes cumbersome) to more complicated systems with more than two dispersive curves.

\MDB{A very first sketch of this model (in its linear version, namely when the springs satisfy Hooke's law) was provided by Lord Kelvin in 1889 to devise a theory of dispersion \cite{thompson1889popular}. Kelvin's model consists of a one-dimensional chain of particles of mass $m_q$ connected by equal springs of elastic constant $\chi_q$, such that each of these masses is attached to another particle of mass $m_r$ by a spring of elastic constant $\chi_r$ (see figure \ref{fig:system} below). This model is explicitly remarked on page 11 of Brillouin's book \cite{brillouin1953wave}, where the dispersion relation is also shown: it consists of two pass-band branches, the low-frequency one being known as the `acoustical branch' and the high-frequency one being known as the `optical branch'. There is a gap of forbidden frequencies between these branches (see figure \ref{fig:disp_rel} below). The existence of branches and gaps is ubiquitous in this type of models. Brillouin's book (op.~cit.) provides an early account of this and many other models, such as the diatomic chain, and constitutes a pivot point in the history of studies on the structure of one-dimensional lattices, where a particular focus was given to understanding how energy propagates in crystals (see also \cite{fermi1955studies, kittle1996}). Interestingly, despite the fact that Brillouin's book is cited by over four thousand publications (most of which deal with metamaterials), and despite the fact that Brillouin calls this model ``Kelvin's model'', Kelvin's model is rarely called by its name in these publications, and it is usually called with generic names such as ``mass-in-mass system''. Following the very recent and relevant paper regarding the linear problem \cite{li2023effective} (which also contains a very good bibliography), we call this model ``Kelvin lattice''. See also \cite{gantzounis2013granular, cveticanin2016theory, nassar2017non, vo2022reinvestigation}.}

\MDB{We are interested in the nonlinear quadratic version of the Kelvin lattice, namely the case when the springs are nonlinear such that the force contains a small extra quadratic dependence on the displacement: $F=\chi \Delta x + \alpha (\Delta x)^2$. Such a case is reminiscent of the celebrated Fermi-Pasta-Ulam-Tsingou lattice \cite{fermi1955studies} ($\alpha$-FPUT for short), which corresponds to our nonlinear quadratic Kelvin lattice in the case when the resonators (of masses $m_r$) are not considered in the picture. The numerical simulations shown in \cite{fermi1955studies} displayed the phenomenon of recurrence: an ordered and reversible nonlinear dynamical exchange of energy amongst the modes of oscillation. The subsequent research on this phenomenon led to the discovery of solitons and integrable nonlinear partial differential equations \cite{zabusky1965interaction}. The recurrence phenomenon led to the apparent paradox that modes of oscillation would not reach equipartition even though they interact nonlinearly due to the $\alpha$-term. Further research on this paradox led to various approaches to explain how energy equipartition is eventually reached at very late times. The interested reader is directed to these reviews on the subject: \cite{onorato2023wave, ford1992fermi, weissert2012genesis, berman2005fermi, carati2005fermi}.}
   
\MDB{Going back to the nonlinear Kelvin lattice and other nonlinear metamaterial models, one could expect the FPUT recurrence phenomenon to be an exception rather than the rule. General nonlinear systems usually display resonances, which by definition break integrability. Although this paper deals with discrete lattices and finite-amplitude effects, where the time scales of the different variables are not necessarily widely separated, it is worth mentioning the strategies related to asymptotic methods that exist in the literature. For example, assuming separation of time scales allows for  perturbation-expansion solutions and nonlinear corrections to the dispersion relation \cite{chen2021wave, settimi2021nonlinear, shen2022nonlinear}. As another example, taking the continuum limit (so the chain is now a one-dimensional continuum) and assuming separation of spatial scales allows for the derivation of nonlinear partial differential equations, a work initiated in 1965 by Zabusky and Kruskal \cite{zabusky1965interaction} with their celebrated discovery of solitons in the Korteweg-de Vries equation.}

This paper is organized as follows:  Sec.~\ref{sec:model} introduces the Kelvin lattice and the procedure of diagonalization of the linear problem. Sec.~\ref{sec:NL} discusses the natural extension to the nonlinear Kelvin lattice, where, besides the theoretical approach, numerical simulations are used to support our analytical findings \MDB{on the effect that three-wave and four-wave resonances have on the system. Sec.~\ref{sec:NKL_FPUT} discusses an assessment of recurrence behaviour of the nonlinear Kelvin lattice, inspired by the well-known FPUT recurrence \cite{fermi1955studies}.} Finally, Sec.~\ref{sec:conc} provides discussions and conclusions. \MDB{Appendix \ref{sec:appendix} contains a derivation of the long-wave continuum limit of the Kelvin lattice.}

\section{The linear Kelvin lattice \label{sec:model}}

\begin{figure}[htpb!]
	\begin{minipage}[h]{1\linewidth}
		\centering
		\includegraphics[width=.9\textwidth]{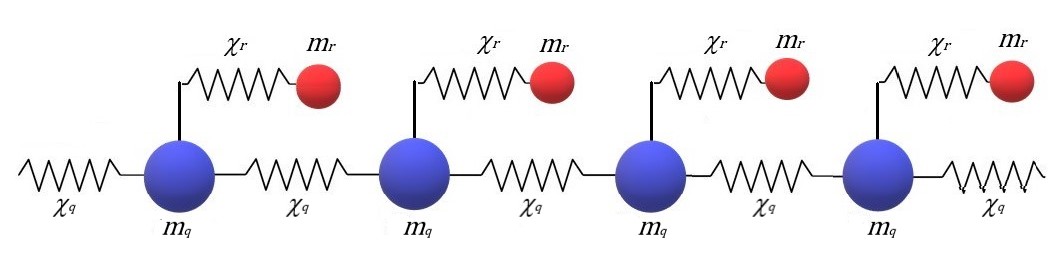}
	\end{minipage}
	\caption{A graphic representation of the Kelvin lattice: a second mass free to oscillate is connected to each mass of the monoatomic chain by means of a spring.}\label{fig:system}
\end{figure}

Let us consider a chain of $ N $ masses, $ m_q $, connected with springs of elastic constant $ \chi_q $. We connect second masses, $ m_r $, to the masses of the chain by means of springs with elastic constant $ \chi_r $ as in Fig. \ref{fig:system}.
The masses $ m_r $ are free to oscillate and are not connected to each other. We assume periodic boundary conditions, {\it \i.e.,} if we denote by $j$ the label of the masses, such that $q_j(t)$ is the displacement of the corresponding mass $m_q$ and $r_j(t)$ is the displacement of the accompanying mass $m_r$, both with respect to their equilibrium positions, then $ q_{N+1}(t)=q_1(t) $ and $ r_{N+1}(t)=r_1(t) $. In this section we consider the simplified case of harmonic potentials, whereby all the springs' potential energies are quadratic in the relative displacements.  This case leads to linear evolution equations which allow us to construct the normal modes and find two branches for the dispersion relation. The case of anharmonic potentials, namely when the potential energies contain higher-order terms, leads to nonlinear interactions between the normal modes and is considered in the next section.

\subsection{\label{sec:eqmotion}Equation of motion for harmonic potentials}

The kinetic energy of the system is given by the sum of the kinetic energies of all particles:
\begin{equation}
	\label{kinetik}
	T=\frac{1}{2}m_q \sum_{j=0}^{N-1} \dot{q}_j^2+
	\frac{1}{2}m_r \sum_{j=0}^{N-1} \dot{r}_j^2,
\end{equation}
while the potential term takes into account the elastic force between the $j$-th mass of the chain with its nearest neighbours and  associated resonator:
\begin{equation}
	\label{potential}
	V=\frac{1}{2} \chi_q \sum_{j=0}^{N-1} (q_{j+1}-q_j)^2 
	+\frac{1}{2} \chi_r \sum_{j=0}^{N-1} (q_j-r_j)^2.
\end{equation}

The Lagrangian of the system is given by $\mathcal{L}=T-V$, then applying the Euler-Lagrange equations 
\begin{equation}
	\label{eulero-lagrange}
	\frac{d}{dt} \bigg( \frac{\partial\mathcal{L} }{\partial \dot{q}_{j} }\bigg)=
	\frac{\partial\mathcal{L} }{\partial{q}_{j} }, \qquad\qquad
	\frac{d}{dt} \bigg( \frac{\partial\mathcal{L} }{\partial \dot{r}_{j} }\bigg)=\frac{\partial\mathcal{L} }{\partial{r}_{j} }, 
\end{equation}
the equations of motion follow: 
\begin{subequations}
	\label{eq_motion}
	\begin{align}
		\ddot{q}_{j}&=\frac{\chi_q}{m_q}(q_{j+1}+q_{j-1}-2q_{j})
		+\frac{\chi_r}{m_q}(r_j-q_j),
		\label{eq_motion1}\\
		\ddot{r}_{j}&=\frac{\chi_r}{m_r}(q_{j}-r_{j}). \label{eq_motion2}
	\end{align}
\end{subequations}

\subsection{\label{sec:motf}Equation of motion in Fourier space}

To decouple Eq.\eqref{eq_motion}, the following Discrete Fourier transforms (DFT) are applied:
\begin{equation}
	\label{DFT_inv} 
		q_{j} = \frac{1}{\sqrt{N}}\sum_{k=0}^{N-1}
		Q_k e^{i\frac{2\pi k}{N}j},\;\;\;\;
				r_{j} = \frac{1}{\sqrt{N}}\sum_{k=0}^{N-1}
		R_k e^{i\frac{2\pi k}{N}j},
\end{equation}

with the completeness of the discrete basis given in terms of the Kronecker delta:
\begin{equation}
	\label{deltaK}
	\delta_{k,k'}=\frac{1}{N} \sum_{j=0}^{N-1} e^{i \frac{2 \pi }{N} (k-k')j}=
	\begin{cases}
		1 & \text{if $k=k'$} \,,\\
		0 & \text{if $k \ne k'$}\,.
	\end{cases}
\end{equation}
Substituting equations \eqref{DFT_inv} and \eqref{deltaK} into equation \eqref{eq_motion} an equivalent system of equations in terms of the Fourier amplitudes is found:

\begin{subequations}
	\label{system}
	\begin{align}
\ddot{Q}_k &=\frac{\chi_r}{m_q}R_k-
\frac{\chi_r+m_q\omega_k^2}{m_q}Q_k, \\
\ddot{R}_k &=\frac{\chi_r}{m_r}(Q_k-R_k),
	\end{align}
\end{subequations}

where $k=0, \ldots, N-1$, and
\begin{equation}
	\label{FPUTomega}
 \omega_k =2 \sqrt{\frac{\chi_q}{m_q}}\Big\vert \sin\Bigl(\frac{\pi k}{N}\Bigr) \Big\vert
\end{equation}

is the dispersion relation of the classical FPUT monoatomic problem, which can be obtained by disconnecting the masses $m_r$ from $m_q$ via setting $\chi_r=0$. 

The system \eqref{system}, for each $k = 0, \ldots, N-1$, consist of independent blocks of two coupled equations. In order to solve these coupled equations, it is useful to consider the Hamiltonian structure. Defining the conjugate momenta
\begin{equation}
	\label{momentum}
	p_j=m_q\dot{q}_j \,,\qquad g_j=m_r\dot{r}_j\,,
\end{equation}
the Hamiltonian is prescribed as follows:
\begin{equation}
	\label{Hamiltoniana}
	H_0= \sum_{j=0}^{N-1}
	\biggl(\frac{p_j^2}{2m_q}+\frac{g_j^2}{2m_r}+
	\frac{1}{2} \chi_q (q_{j+1}-q_j)^2 
	+\frac{1}{2} \chi_r (q_j-r_j)^2 \biggr)\,.
\end{equation}


In terms of the Fourier amplitudes, this Hamiltonian turns out to be
\begin{equation}
	\label{H_Fourier}
	\hat{H}_0 = \sum_{k=0}^{N-1}
	\biggl(\frac{\lvert P_k \rvert^2}{2m_q}+\frac{\lvert G_k \rvert^2}{2m_r} + \frac{1}{2} (\chi_r+m_q\omega_k^2) \lvert Q_k \rvert^2 + 
	\frac{\chi_r}{2} [\lvert R_k \rvert^2-2\Re(Q_k R_k^*)]
	\biggr)
\end{equation}
with
\begin{subequations}
\label{moment_F}
\begin{align}
P_k &= \frac{1}{\sqrt{N}}\sum_{j=0}^{N-1} p_{j} e^{-i\frac{2\pi k}{N}j} =m_q\dot{Q}_k\,,  \\
G_k &= \frac{1}{\sqrt{N}}\sum_{j=0}^{N-1} g_{j} e^{-i\frac{2\pi k}{N}j}=m_r\dot{R}_k\,.
\end{align}
\end{subequations}
The canonical Poisson bracket is obtained:
\begin{equation}
	\label{Poisson}
		\{Q_k,P_{k'}^*\}=\delta_{k,k'}, \qquad\quad \{R_k,G_{k'}^*\}=\delta_{k,k'}
\end{equation}
so the Hamilton equations are
\begin{equation}
	\label{eq_Hamilton}
	\dot{Q_k}=\frac{\partial \hat{H}_0 }{\partial P_k^*}, \qquad 
	\dot{P_k}=-\frac{\partial \hat{H}_0 }{\partial Q_k^*}, \qquad
	\dot{R_k}=\frac{\partial \hat{H}_0 }{\partial G_k^*}, \qquad
	\dot{G_k}=-\frac{\partial \hat{H}_0 }{\partial R_k^*}.
\end{equation}
Note that the displacements are real quantities, thus Fourier amplitudes and momenta are Hermitian, which translate into the inner property $Q_{N-k}=Q_k^*$, $P_{N-k}=P_k^*$, etc. Using equations \eqref{eq_Hamilton} and \eqref{H_Fourier}, the equations of motion \eqref{system} are recovered.

\subsection{\label{sec:Dispersion} The dispersion relation}

For fixed $k = 0, \ldots, N-1$, a more compact formulation of Eqs.\eqref{system} is achieved using a $2\times 2$ matrix form,  

\begin{equation}
	\label{matrixform}
	\ddot{U}=A U\,,
\quad \text{where} \quad
	U=
	\begin{bmatrix}
		Q_k \\ 
		R_k
	\end{bmatrix}
\quad \text{and} \quad
	A=
	\begin{bmatrix}
		-\Bigl(\omega_k^2+\frac{\chi_r}{m_q}\Bigr) & \frac{\chi_r}{m_q} \\ 
		\frac{\chi_r}{m_r} & -\frac{\chi_r}{m_r}
	\end{bmatrix}\,.
\end{equation}
\medskip

\begin{figure}[htpb!]
	\begin{minipage}[h]{1\linewidth}
		\centering
		\includegraphics[width=.5\textwidth]{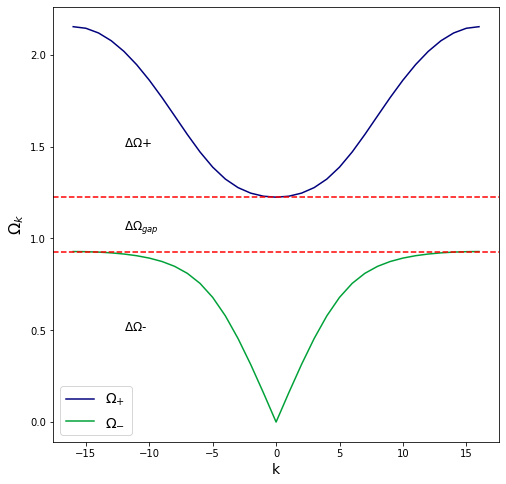}
	\end{minipage}
	\caption{The two branches of the dispersion relation separated by the band gap of forbidden frequencies. We call $+$ the optical branch and $-$ the acoustical branch. The latter contains a degenerate zero-mode, $k=0$, whose frequency is zero.}\label{fig:disp_rel}
\end{figure}

The eigenfrequencies of this system are obtained from the characteristic equation $\det (A+\Omega^2 \mathbb{I})=0$, leading to the diagonalization of $A$ by similarity (see next subsection). In terms of the associated diagonal matrix these eigenfrequencies appear in two branches (i.e. subspaces):
%
\begin{equation}
	\label{Adiag}
	A_D=
	\begin{bmatrix}
		-\Omega_{k(+)}^2 & 0 \\ 
		0 & -\Omega_{k(-)}^2
	\end{bmatrix}\,,
\end{equation}
\smallskip

where
\begin{equation}
	\label{Omega_k}
	\Omega_{k}^\pm=
	\frac{\mathcal{F}_k^+\pm\mathcal{F}_k^-}{2}\,, \qquad \qquad \mathcal{F}^\pm_k
	=\sqrt{\Bigl(\omega_k \pm \sqrt{\frac{\chi_r}{m_r}}\, \Big)^2 + \frac{\chi_r}{m_q}}
\end{equation} 
are the two branches of the dispersion relation, displaying a forbidden frequency band gap (see Fig.~\ref{fig:disp_rel}) of size:
\begin{equation}
	\label{BG}
	\Delta \Omega_{\text{gap}}= \Omega_{\min}^+-\Omega_{\max}^-
	=\Omega_{0}^+-\Omega_{N/2}^-
	=\frac{2\mathcal{F}_0^\pm +\mathcal{F}_{N/2}^-
		- \mathcal{F}_{N/2}^+ }{2}.
\end{equation}

Note that $\mathcal{F}_k^\pm=\Omega_k^+ \pm \Omega_k^-$ and $\Delta \Omega^\pm = \Omega_{N/2}^\pm - \Omega_{0}^\pm$, while characteristic values are
\begin{subequations}
	\label{Omegamaxmin}
	\begin{align}
		\Omega_0^-&=\Omega_{\min}^-=0 \,,\label{A.3a}\\
		\Omega_0^+&=\mathcal{F}_0^\pm=\Omega_{\min}^+=
		\sqrt{\frac{\chi_r}{\bar{\mu}}}\,, \qquad \bar{\mu}=\biggl(\frac{1}{m_q}+\frac{1}{m_r}\biggr)^{-1}\,, \label{A.3b}\\
		\Omega_{N/2}^\pm&=\frac{\mathcal{F}_{N/2}^+\pm\mathcal{F}_{N/2}^-}{2}\,,
		\qquad \mathcal{F}_{N/2}^\pm=
		\sqrt{\bigg(2\sqrt{\frac{\chi_q}{m_q}} \pm \sqrt{\frac{\chi_r}{m_r}}\bigg)^2 + \frac{\chi_r}{m_q}}\,,
		 \label{A.3c}
	\end{align}
\end{subequations} 

so $\Delta \Omega^+ + \Delta \Omega^- + \Delta \Omega_{\text{gap}} = \Omega_{\max}^+$. By varying the parameters $\chi_q, \chi_r, m_q, m_r$ the quantities Eq.\eqref{BG} and
Eq.\eqref{Omegamaxmin} change too, and so do the shapes of the curves in Fig.~\ref{fig:disp_rel}.

\subsection{\label{sec:Diagonal} Diagonalization}

The matrix $A$ is diagonalizable by similarity. Namely, there exists an invertible matrix $P$ such that $A$ is diagonalized: $P^{-1} A P = A_D$, where $A_D$ is diagonal, cf. equation \eqref{Adiag}. 
Such a transformation then diagonalizes Eq.\eqref{matrixform}. Let us denote by $s=\pm$ the two branches. The first step is to solve $A \vec{u}_s=-\Omega_{k(s)}^2\vec{u}_s$, which
returns two eigenvectors:
\begin{equation}
	\label{eigenvector}
	\vec{u}_s=
	\begin{pmatrix}
		1  \\
		\beta_k^s 
	\end{pmatrix}\,, \quad \text{where} \quad
	\beta_k^s=
	\frac{\chi_r+m_q(\omega_k^2-\Omega_{k(s)}^2 )}{\chi_r}
	=\frac{\chi_r}{\chi_r-m_r\Omega_{k(s)}^2}\,, \qquad \qquad s=\pm\,.
\end{equation}

Since $\Omega_{k(+)}^2+\Omega_{k(-)}^2=
		\frac{\chi_r}{\bar{\mu}}+\omega_k^2$,
where $\bar{\mu}$ is the reduced mass defined in Eq.\eqref{A.3b},  a useful equality follows:
\begin{equation}
	\label{bprod}
	\beta_k^+\beta_k^-=
	-\frac{m_q}{m_r}\,.
\end{equation}

From Eq.\eqref{eigenvector} the transition matrix is
\begin{equation}
	\label{cambiobase}
	P=
	\begin{bmatrix}
		1 & 1  \\
		\beta_k^+ & \beta_k^-
	\end{bmatrix}\,.
\end{equation}

Thus, by means of Eq.\eqref{bprod}, its inverse is written as
\begin{equation}
	\label{pinverse}
	P^{-1}=
	\begin{bmatrix}
		\frac{m_q}{\mu_k^+} & \frac{\beta_k^+ m_r}{\mu_k^+}  \\
		\frac{m_q}{\mu_k^-} & \frac{\beta_k^- m_r}{\mu_k^-} 
	\end{bmatrix} \,,
\end{equation}
where we have defined the effective masses
\begin{equation}
	\label{muk}
	\mu_k^s=m_q+(\beta_k^s)^2 m_r \,, \qquad \qquad s=\pm\,,
\end{equation}
such that 
\begin{equation}
\label{eq:relations}
\frac{1}{\mu_k^+}+\frac{1}{\mu_k^-}=\frac{1}{m_q}\,, \qquad\quad
\frac{\beta_k^+}{\mu_k^+}+\frac{\beta_k^-}{\mu_k^-}=0\,, \qquad\quad
\frac{(\beta_k^+)^2}{\mu_k^+}+\frac{(\beta_k^-)^2}{\mu_k^-}=\frac{1}{m_r}\,.
\end{equation}

Now from Eqs.\eqref{matrixform}, \eqref{Adiag}, \eqref{cambiobase} and \eqref{pinverse} we get $\ddot{U}=AU=PA_D P^{-1}U$,
from which $P^{-1}\ddot{U}=A_D\, P^{-1}U$, or, defining 
\begin{equation}
	\label{sistem_diag}
	\begin{bmatrix}
		\widetilde{Q}^+_k\\ 
		\widetilde{Q}^-_k
	\end{bmatrix} :=  P^{-1}U = \begin{bmatrix}
		\frac{m_q}{\mu_k^+} & \frac{\beta_k^+ m_r}{\mu_k^+}  \\
		\frac{m_q}{\mu_k^-} & \frac{\beta_k^- m_r}{\mu_k^-} 
	\end{bmatrix}  \begin{bmatrix}
		Q_k \\ 
		R_k
	\end{bmatrix}\,,
\end{equation}
we obtain 	the diagonal system $\frac{d^2}{dt^2}\begin{bmatrix}
		\widetilde{Q}^+_k\\ 
		\widetilde{Q}^-_k
	\end{bmatrix}=A_D\begin{bmatrix}
		\widetilde{Q}^+_k\\ 
		\widetilde{Q}^-_k
	\end{bmatrix}$,
which is a fully decoupled system of two harmonic oscillators:
\begin{equation}
	\label{oscillators}
	\ddot{\widetilde{Q}}{}^s_k+\Omega^2_{k(s)}\widetilde{Q}_k^s=0\,, \qquad \qquad s = \pm\,,
\end{equation}
corresponding to the two branches of the dispersion relation.

The coordinate transformation \eqref{sistem_diag} can be inverted and simplified using equations \eqref{eq:relations} to give
\begin{equation}
	\label{diffeo}
	\begin{cases}
		Q_k=\widetilde{Q}^+_k+\widetilde{Q}^-_k \,,\\
		R_k=\beta_k^+ \widetilde{Q}^+_k+ \beta_k^- \widetilde{Q}^-_k\,.
	\end{cases}
\end{equation}
From Eqs.\eqref{moment_F} and \eqref{diffeo}, using Eqs.\eqref{bprod} and \eqref{muk}, the kinetic term in Eq.\eqref{H_Fourier} becomes
$	\widetilde{T}=\sum_{k=0}^{N-1} \biggl( 
	\frac{\lvert \widetilde{P}^+_k \rvert^2}{2 \mu_k^+} +
	\frac{\lvert \widetilde{P}^-_k \rvert^2}{2\mu_k^-}
	\biggr)
$, 
where 
$	\widetilde{P}^s_k:=\mu_k^s \dot{\widetilde{Q}}{}_k^s   
$. As for the potential term, using the identities
$m_q\omega_k^2+\chi_r(1-\beta_k^s)^2=\mu_k^s\Omega_{k(s)}^2
$ and $\beta_k^+ + \beta_k^-=
	1-\frac{m_q}{m_r}+\frac{m_q\omega_k^2}{\chi_r}$, 
we find
 $	\widetilde{V}= \sum_{k=0}^{N-1}
	\biggl( \frac{1}{2}\mu_k^+\Omega_{k(+)}^2 
	\lvert \widetilde{Q}^+_k \rvert^2 + 
	\frac{1}{2}\mu_k^-\Omega_{k(-)}^2 
	\lvert \widetilde{Q}^-_k \rvert^2 
	\biggr)$.

So, in summary we have the canonical transformation, valid for all $k=0, \ldots, N-1$ and all $s=\pm$:
\begin{equation}
	\label{trasf_canonica}
	\begin{cases}
		\widetilde{Q}^s_k=\frac{m_q}{\mu_k^s}Q_k+
		\beta_k^s\frac{m_r}{\mu_k^s}R_k\,,	\\
		\widetilde{P}^s_k=P_k+\beta_k^s G_k \,,
	\end{cases}
\end{equation}
for which the Hamiltonian Eq.\eqref{H_Fourier} presenting coupled terms becomes the diagonalized quadratic form
\begin{equation}
	\label{H_diag}
	\widetilde{H}_0= \sum_{k=0}^{N-1}\sum_{s=\pm}
	\biggl( 
	\frac{\lvert \widetilde{P}^s_k \rvert^2}{2 \mu_k^s} +
	\frac{1}{2}\mu_k^s\Omega_{k(s)}^2 
	\lvert \widetilde{Q}^s_k \rvert^2 
	\biggr)\,,
\end{equation}
with Poisson brackets 
$		\{\widetilde{Q}_k^s,\widetilde{P}_{k'}^{s*}\}
		=\delta_{k,k'}
$
 and the Hamilton equations
$	\dot{\widetilde{Q}_k^s}=
	\frac{\partial \widetilde{H}_0}{\partial \widetilde{P}_k^{s*}}\,,
 \quad
	\dot{\widetilde{P}_k^s}=
	-\frac{\partial \widetilde{H}_0}{\partial \widetilde{Q}_k^{s*}}
$,
 which are equivalent to Eqs.~\eqref{oscillators}. Note that the acoustical mode 
 corresponding to $(k,s)=(0,-)$ is characterized by only kinetic energy which remains 
 constant in time. 
Therefore, we introduce the \emph{normal variables}  via the canonical transformation: 
\begin{equation}
	\label{normvar}
	\begin{cases}
		a_k^s=\frac{i}{\sqrt{2\mu_k^s\Omega_k^s}} 
		(\widetilde{P}_k^s-i\mu_k^s\Omega_k^s\widetilde{Q}_k^s)\,, \\
		a_{N-k}^{s*}=-\frac{i}{\sqrt{2\mu_k^s\Omega_k^s}}
		(\widetilde{P}_k^s+i\mu_k^s\Omega_k^s\widetilde{Q}_k^s)\,,
	\end{cases} 
	\qquad
	\begin{cases}
		\widetilde{Q}_k^s=\frac{1}{\sqrt{2\mu_k^s\Omega_k^s}}(a_k^s+a_{N-k}^{s*})\,,\\
		\widetilde{P}_k^s=-i \sqrt{\frac{\mu_k^s \Omega_k^s}{2}} (a_k^s-a_{N-k}^{s*})\,,
	\end{cases}
	\qquad \text{where} \quad (k,s) \neq (0,-)  \,.
\end{equation} 
As the above transformation is not applied to the acoustical mode $(k,s)=(0,-)$, the original momentum variable $\widetilde{P}^-_0$ is kept, which is conserved in time by virtue of Hamilton equations, as $\widetilde{Q}^-_0$ is ignorable. For the normal variables the system \eqref{oscillators} becomes decoupled and reads, along with its solution, 
\begin{equation}
	\label{universal_lin_sol}
	i\frac{da_k^s}{dt}=\Omega_k^s a_k^s \quad \Longrightarrow \quad 	a_k^s(t)=a_k^s(0)e^{-i \Omega_k^s t}\,, \quad (k,s) \neq (0,-)\,.
\end{equation}
The above solution implies 
$	\lvert a_k^s(t) \rvert^2  = \lvert a_k^s(0) \rvert^2\,, \,\,  t \in \mathbb{R}$,
namely, the wave-action at each normal mode does not evolve in time. In the normal variables, the Hamiltonian Eq.\eqref{H_diag} takes the form: 
\begin{equation}
	\label{normalH}
	\bar{H}_0= \frac{\lvert \widetilde{P}^-_0 \rvert^2}{2 \mu_0^-}+\Omega_0^+ \lvert a_0^+ \rvert^2 + \sum_{k=1}^{N-1} \bigl( \Omega_k^+ \lvert a_k^+ \rvert^2 + 
	\Omega_k^- \lvert a_k^- \rvert^2 \bigr)  ,
\end{equation}
and $
	\mathcal{E}_k^s=\Omega_k^s \lvert a_k^s(t) \rvert^2$
is the energy associated with the $s$-branch $k$-th mode, which does not evolve in time.

\subsection{\label{sec:eqmotsol} Analytical solutions to the equations of motion}
\subsubsection{Periodic boundary conditions}
Using equation \eqref{universal_lin_sol} and recalling the identities  $\beta_k^s=\beta_{N-k}^s $, $\Omega^s_{k}=\Omega^s_{N-k}$,  $\mu_k^s=\mu_{N-k}^s$, $\widetilde{Q}^s_k=\widetilde{Q}^{s*}_{N-k}$, the solution for $\widetilde{Q}^s_k$ can be written in the form:
\begin{equation}
	\label{solution}
	\widetilde{Q}^s_k=\frac{1}{2}
	(A_k^s e^{-i\Omega_k^st}+A_{N-k}^{s*}e^{i\Omega_k^st}),
	\quad A_N^+ = A_0^+, \quad  A_k^s \in \mathbb{C},
	\qquad k=0, \ldots, N-1, \qquad s=\pm, \qquad (k,s) \neq (0,-)\,,
\end{equation}
while the ignorable coordinate evolves as $\widetilde{Q}^-_0(t) = C_1 + C_2 \,t$, with constants $C_1=\widetilde{Q}^-_0(0), C_2 = \frac{\widetilde{P}^-_0}{2 \mu_0^-}$. From here on we set $C_1=C_2=0$, obtained after an appropriate Galilean transformation. In the original variables, using Eq.\eqref{diffeo}, the solution in Fourier space reduces to:
\begin{subequations}
	\label{solution2}
	\begin{align}
	Q_k&=\frac{1}{2}  \sum_{s=\pm}
	(A_k^s e^{-i\Omega_k^st}+A_{N-k}^{s*}e^{i\Omega_k^st}),  \\
	R_k&=\frac{1}{2} \sum_{s=\pm}
	\beta_k^s(A_k^s e^{-i\Omega_k^st}+A_{N-k}^{s*}e^{i\Omega_k^st}),
	\end{align}
\end{subequations}
and in physical space, denoting $A_k^s=\sqrt{N} B_k^s  e^{i\varphi_k^s}$ with $0\leq B_k^s$ and $0\leq \varphi_k^s < 2\pi$, it reads: 
\begin{subequations}
	\label{solution3}
	\begin{align}
	q_{j}(t) &= \frac{1}{2}\sum_{k=1}^{N-1}\sum_{s=\pm} \bigl(
	B_k^s  
	e^{i(\frac{2\pi k}{N}j-\Omega_k^s t+\varphi_k^s)}+
	 B_{N-k}^s  
	e^{i(\frac{2\pi k}{N}j+\Omega_k^s t-\varphi_{N-k}^s)} \bigr) + \frac{B_0^+}{2} \bigl(
	e^{i(-\Omega_0^+ t+\varphi_0^+)}+
	e^{i(\Omega_0^+ t-\varphi_0^+)} \bigr)\,,\\
		r_{j}(t) &= \frac{1}{2}\sum_{k=1}^{N-1}\sum_{s=\pm} \beta_k^s \bigl(
	B_k^s  
	e^{i(\frac{2\pi k}{N}j-\Omega_k^s t+\varphi_k^s)}+
	B_{N-k}^s  
	e^{i(\frac{2\pi k}{N}j+\Omega_k^s t-\varphi_{N-k}^s)} \bigr)+ \frac{\beta_0^+ B_0^+}{2} \bigl(
	e^{i(-\Omega_0^+ t+\varphi_0^+)}+
	e^{i(\Omega_0^+ t-\varphi_0^+)} \bigr)\,.
	\end{align}
\end{subequations}

Rearranging the symmetric contributions from $k$ and $N-k$ in the sums above, we find the solution as a superposition
$$		q_{j}(t) = \sum_{(k,s)\neq (0,-)} B_k^s
		\cos\Bigl(\frac{2\pi k}{N}j-\Omega_k^s t+\varphi_k^s\Bigr)\,,\qquad 
		r_{j}(t) = \sum_{(k,s)\neq (0,-)} \beta_k^s B_k^s \cos\Bigl(\frac{2\pi k}{N}j-\Omega_k^s t+\varphi_k^s\Bigr)\,,$$
where real amplitudes $B_k^s$ and phases $\varphi_k^s$ are determined by the initial conditions.

\subsubsection{Fixed boundary conditions}

Consider a chain composed by $N+1$ masses and as many resonators. Fixed boundary conditions implies that $q_j(t)=r_j(t)=0, \forall t \in \mathbb{R}^+$, for $j=0,N$. In this case a real Fourier transform has to be used to write the equation of motion in the Fourier space
\begin{equation}
	\label{sintrans}
	q_j=\sum_{k=1}^{N-1} Q_k \sin \Bigl( \frac{ \pi k}{N} j \Bigr) \qquad\quad r_j=\sum_{k=1}^{N-1} R_k \sin \Bigl( \frac{ \pi k}{N} j \Bigr)
\end{equation}

to get the solutions
\begin{subequations}
	\label{fixBC_sol}
	\begin{align}
		q_j(t)&=\sum_{k=1}^{N-1}\sum_{s=\pm} [A_k^s\cos(\bar{\Omega}_k^s t)+B_k^s\sin(\bar{\Omega}_k^s t) ]\sin \Bigl( \frac{ \pi k}{N} j \Bigr)
		\\
		r_j(t)&=\sum_{k=1}^{N-1}\sum_{s=\pm} \bar{\beta}_k^s [A_k^s\cos(\bar{\Omega}_k^s t)+B_k^s\sin(\bar{\Omega}_k^s t) ]\sin \Bigl( \frac{ \pi k}{N} j \Bigr)
	\end{align}
\end{subequations}

where $\bar{\Omega}_k^s$ and then $\bar{\beta}_k^s$ have the same form of Eqs.~\eqref{Omega_k}, \eqref{eigenvector}, but with 
\begin{equation}
	\label{FPUTomega_fix}
	\bar{\omega}_k =2 \sqrt{\frac{\chi_q}{m_q}}\Big\vert \sin\Bigl(\frac{\pi k}{2N}\Bigr) \Big\vert\,.
\end{equation}

The fact that Equations \eqref{fixBC_sol} are a superposition of stationary waves is more clearly visible imposing null initial velocities, so that all $B_k^s$ are zero, and an initial sinusoidal pattern $q_j(0)=r_j(0)=\sin(n\pi j/N)$ with $0<n<N, n \in \mathbb{N}$. In this case only the $n$-th harmonic is involved and each of Eq.s\eqref{fixBC_sol} is made by the sum of four waves, two by two of the same amplitude given by the half of
$	A_n^\pm=\frac{m_r m_q }{\chi_r \mu_n^\pm} \bar{\Omega}_{n(\mp)}^2  $ and $
	\bar{\beta}_n^\pm A_n^\pm$,
and traveling in opposite directions, that is
\begin{subequations}
	\label{oppwavesin}
	\begin{align}
		q_j(t)&=\sum_{s=\pm} \frac{A_n^s}{2}
		\Bigl[
		\sin \Bigl(  \frac{ n \pi}{N} j  - \bar{\Omega}_n^s t \Bigr)+
		\sin \Bigl( \frac{ n \pi}{N} j + \bar{\Omega}_n^s t \Bigr)
		\Bigr]\,,
		\\
		r_j(t)&=\sum_{s=\pm} \frac{\bar{\beta}_n^s A_n^s}{2}
		\Bigl[
		\sin \Bigl(  \frac{ n \pi}{N} j  - \bar{\Omega}_n^s t \Bigr)+
		\sin \Bigl( \frac{ n \pi}{N} j + \bar{\Omega}_n^s t \Bigr)
		\Bigr]\,.
	\end{align}
\end{subequations}

Equating to zero Eqs.~\eqref{oppwavesin}, the value
\begin{equation}
	j_l^*=
	\begin{cases}
		l\frac{N}{n}  \qquad l=0,\dots,n \quad \text{if} \quad n \vert N \\
		lN \qquad l=0,1 \qquad\quad \text{otherwise}
	\end{cases}
\end{equation}

is found. This means that, if $n$ is a divisor of $N$, the stationary waves present $n+1$ nodes for discrete values of $j$, i.e. the $j_l^*$-th masses and resonators stay still during the time evolution of the system. Otherwise, the chain has only the two trivial nodes at the ends. 

\section{\label{sec:NL} The nonlinear Kelvin lattice }

\subsection{\label{sec:EqmotNL} Equations of motion }

An anharmonic potential can be represented via a cubic power term { in the Hamiltonian, which leads to three-wave interaction systems.  This kind of systems has been widely studied in the past, see \cite{kaup1979space,craik1988wave}; the complication here is that, as will be shown below, nonlinear interactions may take place between the two branches of the dispersion relation.
}
Adding the nonlinear contribution to our system gives the following  Hamiltonian:
\begin{equation}
	\label{Hamiltoniana_NL}
		H=H_0 +\frac{\alpha}{3} \sum_{j=0}^{N-1}
		[(q_{j+1}-q_j)^3 + (q_j-r_j)^3] \,,
\end{equation}

where $H_0$ is given by Eq.~\eqref{Hamiltoniana}. From this, the equations of motion are obtained:
\begin{subequations}
	\label{eq_motion_nl}
	\begin{align}
		\ddot{q}_{j}&=\frac{\chi_q}{m_q}(q_{j+1}+q_{j-1}-2q_{j})
		+\frac{\chi_r}{m_q}(r_j-q_j)+\frac{\alpha}{m_q}
		[(q_{j+1}-q_{j})^2-(q_{j}-q_{j-1})^2-(q_{j}-r_{j})^2]
		\label{eq_motion1nl}\,,\\
		\ddot{r}_{j}&=\frac{\chi_r}{m_r}(q_{j}-r_{j})+
		\frac{\alpha}{m_r}(q_{j}-r_{j})^2
		\label{eq_motion2nl}\,.
	\end{align}
\end{subequations}

\subsection{\label{sec:FourierNL} Fourier space }

Following similar steps to the linear case, the DFT \eqref{DFT_inv} is applied to Eq.~\eqref{eq_motion_nl}, giving
\begin{subequations}
	\label{motionF_NLQR}
	\begin{align}
		\ddot{Q}_{1}&=
		\frac{\chi_r}{m_q}R_{1}-\frac{\chi_r+m_q\omega_k^2}{m_q}Q_{1}
		-\frac{1}{\sqrt{N}}\frac{\alpha}{m_q} \sum_{k_2,k_3}
		[S_{-1,2,3}Q_{2}Q_{3}-2Q_{2} R_{3}+R_{2}R_{3}]\delta_{1,2+3}
		\label{motionF_NLQ}\,,\\
		\ddot{R}_1 &=\frac{\chi_r}{m_r}(Q_1-R_1)+
		\frac{1}{\sqrt{N}}\frac{\alpha}{m_r} \sum_{k_2,k_3}
		[Q_2Q_3-2Q_2R_3+R_2R_3]\delta_{1,2+3}
		\label{motionF_NLR}\,,
	\end{align}
\end{subequations}
where $Q_i$ ($R_i$) is shorthand for $Q_{k_i}$ ($R_{k_i}$),  the Kronecker delta now accounts for  Umklapp interactions:
\begin{equation}
	\label{delta1,2+3}
	\delta_{1,2+3}=
	\begin{cases}
		1 & \text{if $k_1=k_2+k_3 \pmod N$}\,, \\
		0 & \text{otherwise}\,,
	\end{cases}
\end{equation}
and
\begin{equation}
	\label{S}
	S_{\pm1,2,3}=1\pm 8 \,i\,\mathrm{e}^{i(\pm k_1+k_2+k_3)\frac{\pi}{N} }
	\sin\Bigl(\frac{\pi k_1}{N}\Bigr)
	\sin\Bigl(\frac{\pi k_2}{N}\Bigr)
	\sin\Bigl(\frac{\pi k_3}{N}\Bigr).
\end{equation}
The adding term 1 is related to the presence of the resonators, while the one given by the product of 
trigonometric functions is the same as the one derived for the monoatomic 
chain, see \cite{bustamante2019exact}.
Eqs.~\eqref{motionF_NLQR} can be equivalently obtained 
from Eq.~\eqref{eq_Hamilton} using the following Hamiltonian:
\begin{equation}
		\label{H_Fourier_NL_S}
		\hat{H}=\hat{H}_0 
		+\frac{\alpha}{3\sqrt{N}}\sum_{1,2,3} 
		[S_{1,2,3}^* Q_1Q_2Q_3 -R_1R_2R_3 
		+3Q_1R_2R_3-3Q_1Q_2R_3] \delta_{-1,2+3}\,,
\end{equation}
where $\hat{H}_0$ is given by Eq.~(\eqref{H_Fourier}). Notice that $S_{1,2,3}$ is invariant under permutations of $k_1,k_2,k_3$, corresponding to a symmetry over the exchange of all wavenumbers. 
Now, by introducing the variable $\widetilde{Q}^s_k$ from equation \eqref{trasf_canonica}, we obtain equivalent equations of motion:
\begin{equation}
	\label{systemNL_diag}
	\begin{split}
		\ddot{\widetilde{Q}}_1^++\Omega_{1(+)}^2\widetilde{Q}_1^+ &=
		\frac{1}{\sqrt{N}}\frac{\alpha}{\mu_1^+}
		\sum_{2,3} \{
		V_{-1,2,3}^{+++}\widetilde{Q}^+_2\widetilde{Q}^+_3+
		2V_{-1,2,3}^{++-}\widetilde{Q}^+_2\widetilde{Q}^-_3+
		V_{-1,2,3}^{+--}\widetilde{Q}^-_2\widetilde{Q}^-_3\}\delta_{1,2+3},\\
		\ddot{\widetilde{Q}}_1^-+\Omega_{1(-)}^2\widetilde{Q}_1^- &=
		\frac{1}{\sqrt{N}}\frac{\alpha}{\mu_1^-}
		\sum_{2,3} \{
		V_{-1,2,3}^{-++}\widetilde{Q}^+_2\widetilde{Q}^+_3+
		2V_{-1,2,3}^{-+-}\widetilde{Q}^+_2\widetilde{Q}^-_3+
		V_{-1,2,3}^{---}\widetilde{Q}^-_2\widetilde{Q}^-_3\}\delta_{1,2+3},
	\end{split}
\end{equation}
where the coefficients are 
\begin{equation}
		V_{1,2,3}^{s_1s_2s_3}=\beta_1^{s_1} + \beta_2^{s_2} + \beta_3^{s_3} -\beta_1^{s_1}\beta_2^{s_2} -\beta_1^{s_1}\beta_3^{s_3} 
		-\beta_2^{s_2}\beta_3^{s_3} + \beta_1^{s_1}\beta_2^{s_2}\beta_3^{s_3} - S_{123}\,.
\end{equation}

Unlike the linear case, see eq.\eqref{oscillators}, the oscillators are now forced by interactions terms between modes of the two branches of the dispersion relation. In the new variables, the Hamiltonian reads:
\begin{equation}
	\begin{split}
		\widetilde{H}=\widetilde{H}_0
		-\frac{\alpha}{3\sqrt{N}} \sum_{1,2,3}
		\{&V_{1,2,3}^{+++*}
		\widetilde{Q}^+_1\widetilde{Q}^+_2\widetilde{Q}^+_3 +
		V_{1,2,3}^{---*}
		\widetilde{Q}^-_1\widetilde{Q}^-_2\widetilde{Q}^-_3 + \\
		+3 &V_{1,2,3}^{++-*}
		\widetilde{Q}^+_1\widetilde{Q}^+_2\widetilde{Q}^-_3 +
		3V_{1,2,3}^{+--*}
		\widetilde{Q}^+_1\widetilde{Q}^-_2\widetilde{Q}^-_3
		\} \delta_{-1,2+3}.
	\end{split}
\end{equation}
The last step consists of writing the equation in normal variables, using the transformations in \eqref{normvar}:
\begin{equation}
\begin{split}
	\label{universal}
	i\frac{da_1^s}{dt}=\Omega_1^s a_1^s + \sum_{2,3}
	\{ [\, &\bar{V}_{-1,2,3}^{s++}\, a_2^+ a_3^+ + 
	\bar{V}_{-1,2,3}^{s+-}\, a_2^+a_3^- +
	\bar{V}_{-1,2,3}^{s--} \, a_2^- a_3^- \,] \, \delta_{1,2+3} +  \\
	+2[\,&\bar{V}_{1,-2,3}^{s++*}\, a_2^+ a_3^{+*} + \bar{V}_{1,-2,3}^{s+-*}\, a_2^+ a_3^{-*} 
	+\bar{V}_{1,3,-2}^{s+-*}\, a_2^- a_3^{+*} +
	\bar{V}_{1,-2,3}^{s--*}\, a_2^- a_3^{-*} \,]\, \delta_{1,2-3} +\\
	+[\,&\bar{V}_{1,2,3}^{s++}\, a_2^{+*} a_3^{+*} + 
	2\bar{V}_{1,2,3}^{s+-}\, a_2^{+*} a_3^{-*} +
	\bar{V}_{1,2,3}^{s--}\, a_2^{-*} a_3^{-*}\,]\, \delta_{-1,2+3}\},
\end{split}
\end{equation}
where
\begin{equation}
	\label{unicoeff}
	\bar{V}_{1,2,3}^{s_1s_2s_3}= \gamma_{1,2,3}^{s_1s_2s_3} V_{1,2,3}^{s_1s_2s_3},\qquad \qquad 	\gamma_{1,2,3}^{s_1s_2s_3}=
	-\frac{\alpha}{2\sqrt{2N\mu_1^{s_1}\mu_2^{s_2}\mu_3^{s_3}
			\Omega_1^{s_1}\Omega_2^{s_2}\Omega_3^{s_3}}}.
\end{equation}

\subsection{Resonances}
The system of equations in \eqref{universal} indicates that there are different types of three-wave  interactions that can take place within the same branch of the dispersion relation or even between the two branches of the dispersion relation. By construction, transfer of energy between normal modes takes place only if at least one of the Kronecker deltas is satisfied. Among all these interactions, the most relevant ones are those that satisfy the resonant condition, {i.e.}, those for which, besides the condition on wavenumbers, an analogous condition on frequencies is satisfied.  For $N\rightarrow \infty$, keeping the distance between the masses fixed, wavenumbers can be treated as continuous variables and three-wave exact resonances  exist. For example, if we set $m_q=m_r=\chi_q=\chi_r=1$, the  resonant interactions between two acoustical and one optical mode can take place if one of the following conditions is satisfied:
\begin{equation}
\begin{split}
	\label{resonance}
\left(k_1=k_2+k_3,	\qquad 		\Omega_1^{+}=\Omega_2^{-}+\Omega_3^{-} \right)\qquad {\rm or} \qquad 
\left(k_1=k_2-k_3,	\qquad 		\Omega_1^{-}=\Omega_2^{+}-\Omega_3^{-}\right).\\
\end{split}
\end{equation}
These interactions correspond to the terms $\bar{V}_{-1,2,3}^{+--}\, a_2^-a_3^-\delta_{1,2+3}$ and  
$\bar{V}_{1,-2,3}^{-+-}\, a_2^+a_3^{-*}\delta_{1,2-3}$ in equations \eqref{universal}. 
For finite $N$, resonant triads exist for specific values of the parameters $m_q$, $m_r$, $\chi_q$ and $\chi_r$. 
In general, non resonant interactions may contribute to higher order resonances. Indeed, there is a well-known asymptotic procedure in analytical mechanics, known as {\it near identity transformation} or {\it Lie transform} that allows one to remove analytically the non resonant terms; for Hamiltonian systems, the transformation can be canonical \cite{krasitskii1990canonical}. The non resonant terms contribute to higher order interactions which may or may not satisfy the resonant condition. A typical example where the technique is employed is the water wave problem where, because of the shape of the dispersion relation, three-wave resonances do not exists and a canonical transformation is employed to remove them and recast them in terms of cubic nonlinearity, which results in four-wave interactions \cite{krasitskii1994reduced}. Without going into the details, we mention that, in the model here discussed, four-wave resonant interactions may naturally appear on a  time scale larger than the triad time scale. For discrete systems, resonant quartets may be of the form:
\begin{equation}
\begin{split}
	\label{4resonance}
(k_1+k_2=k_3+k_4,	\qquad 		\Omega_1^{+}+\Omega_2^{-}=\Omega_3^{+}+\Omega_4^{-}) \qquad {\rm or} \qquad 
(k_1+k_2=k_3+k_4,	\qquad 		\Omega_1^{\pm}+\Omega_2^{\pm}=\Omega_3^{\pm}+\Omega_4^{\pm}),\\
      \end{split}
\end{equation}
\begin{equation}
\begin{split}
	\label{4resonance_s}
\text{with} \qquad (k_1,k_2,k_3,k_4)=\left(n,\frac{N}{2}-n,N-n,\frac{N}{2}+n\right), \qquad \text{for \,\, some \,\,} 1\leq n \leq N-1\,.
      \end{split}
\end{equation}
See \cite{bustamante2019exact,pistone2019universal}  for details.
Numerical simulations of the original equation of motion, displaying three- and four-wave resonant interactions, will be shown in the next section. 

\MDB{We remark that for the standard FPUT lattice (regardless of the boundary conditions), three-wave resonant interactions are forbidden and the lowest order of resonant interactions is the four-wave one, observed in the case of periodic boundary conditions (see \cite{rink2001symmetry, onorato2015route, bustamante2019exact} for details). The introduction of the resonators allows for a transfer of energy that takes place on a faster time scale with respect to the standard FPUT one. Moreover, we also recall that from the FPUT lattice, in the limit of large $N$ and small distances between the masses, the KdV equation is recovered. In the Appendix, we report a similar calculation and find that, in the continuum limit, the nonlinear quadratic Kelvin lattice reduces to a Boussinesq equation coupled with a continuum of harmonic oscillators.}

\subsection{\label{sec:Numsim} Numerical simulations}
To support our analytical findings, we perform numerical simulations of the equations of motion \eqref{eq_motion_nl}. We use a symplectic integrator scheme \cite{yoshida1990construction} for the time-marching,   ensuring an error around the eleventh digit in the Hamiltonian over our integration time.

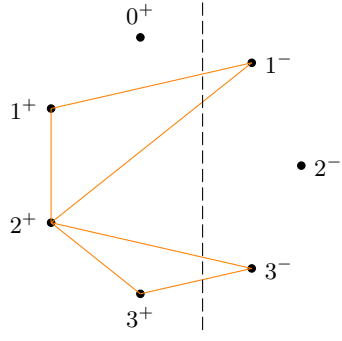
\begin{figure}[H]
    \centering
\begin{tikzpicture}
   \newdimen\R
   \R=1.75cm
   \draw(0:\R)[white] \foreach \x in {51.43,102.86,154.29,205.71,257.14,308.57,360} {  -- (\x:\R) };
   \foreach \x/\l/\p in
     { 51.43/{$1^-$}/right, 
      102.86/{$0^+$}/above, 
      154.29/{$1^+$}/left, 
      205.71/{$2^+$}/left, 
      257.14/{$3^+$}/below, 
      308.57/{$3^-$}/right, 
      360/{$2^-$}/right 
     }
     \node[inner sep=1pt,circle,draw,fill,label={\p:\l}] at (\x:\R) {};
     \draw
    [orange] (-1.5767523096670022, 0.7591786047826782) -- ( 1.0910730390797219, 1.3682322987685023); 
     \draw
    [orange] (-1.5767523096670022, 0.7591786047826782) -- (-1.5767523096670024,-0.7591786047826777); 
    \draw
    [orange] ( 1.0910730390797219, 1.3682322987685023) -- (-1.5767523096670024,-0.7591786047826777); 
    \draw
    [orange] (-0.3894967124495318, -1.7061044255821525) -- (-1.5767523096670024,-0.7591786047826777); 
    \draw
    [orange] (1.0910730390797214, -1.3682322987685025) -- (-1.5767523096670024,-0.7591786047826777); 
    \draw
    [orange] (1.0910730390797214, -1.3682322987685025) -- (-0.3894967124495318, -1.7061044255821525); 

    \draw
    [black,dash pattern= on 5pt off 2pt] (0.25*\R,- 1.25*\R) -- (0.25*\R,1.25*\R);
    \end{tikzpicture}
        \caption{ A visual scheme of the modes interactions.  The modes are labelled $k^s$, where $k=0,1,2,3$ denotes the wavenumber and $s=\pm$ denotes the branch ($-$: acoustical, $+$: optical). The orange continuous line connects two resonant triads. The vertical dashed line separates the frequencies belonging to the optical and acoustical modes.}
        \label{fig:heptagon}
\end{figure}
For simplicity, we consider the case $N=4$, $\alpha=0.2$. All other parameters are fixed to one, except for $\chi_r$ which is selected in such a way to allow exact three-wave resonances. Eight normal modes are accessible: four belonging to the optical branch and four to the acoustical branch; however, the $k=0$ acoustical mode remains constant in time. Therefore, only seven modes are active. These are shown schematically in Fig.~\ref{fig:heptagon}. For each of the two triads $\lbrace (1^-, 1^+, 2^+),(3^-, 3^+, 2^+) \rbrace$, the quantity $\Omega_{k_1}^- + \Omega_{k_2}^+ - \Omega_{k_3}^+ $ is equal to  zero if $\chi_r$ satisfies the following equation:
\begin{equation}
    \label{eq:chi}
    2 \chi_r^2 + \sqrt2 \chi_r^{3/2}  = 1.
\end{equation}
The approximated value 
$\chi_r \approx 0.3522011287389576$ is the solution to Eq.\eqref{eq:chi} that will be used in simulations. 
Initial conditions are supplied by prescribing  the value of $a_k^{s}$ for some specific modes related to the chosen resonant triad. More specifically, the initial conditions satisfy the following relations:
\begin{equation}
\label{eq:IC_reso}
\begin{split}
 &\mathcal{E}_k^+(t=0)=\mathcal{E}_k^-(t=0)=0.5 \qquad {\rm for} \qquad k=1, \\
&  \mathcal{E}_k^+(t=0)=\mathcal{E}_k^-(t=0)=0 \,\,\,\,\, \qquad {\rm for}\qquad k \ne 1;
 \end{split}
 \end{equation}
 therefore, at $t=0$, only normal modes $\lbrace 1^-, 1^+ \rbrace$ have energy.
\begin{figure}[htpb!]
	\centering
	\includegraphics[scale=0.26,trim={2.9cm 1cm 4cm 1cm},clip]{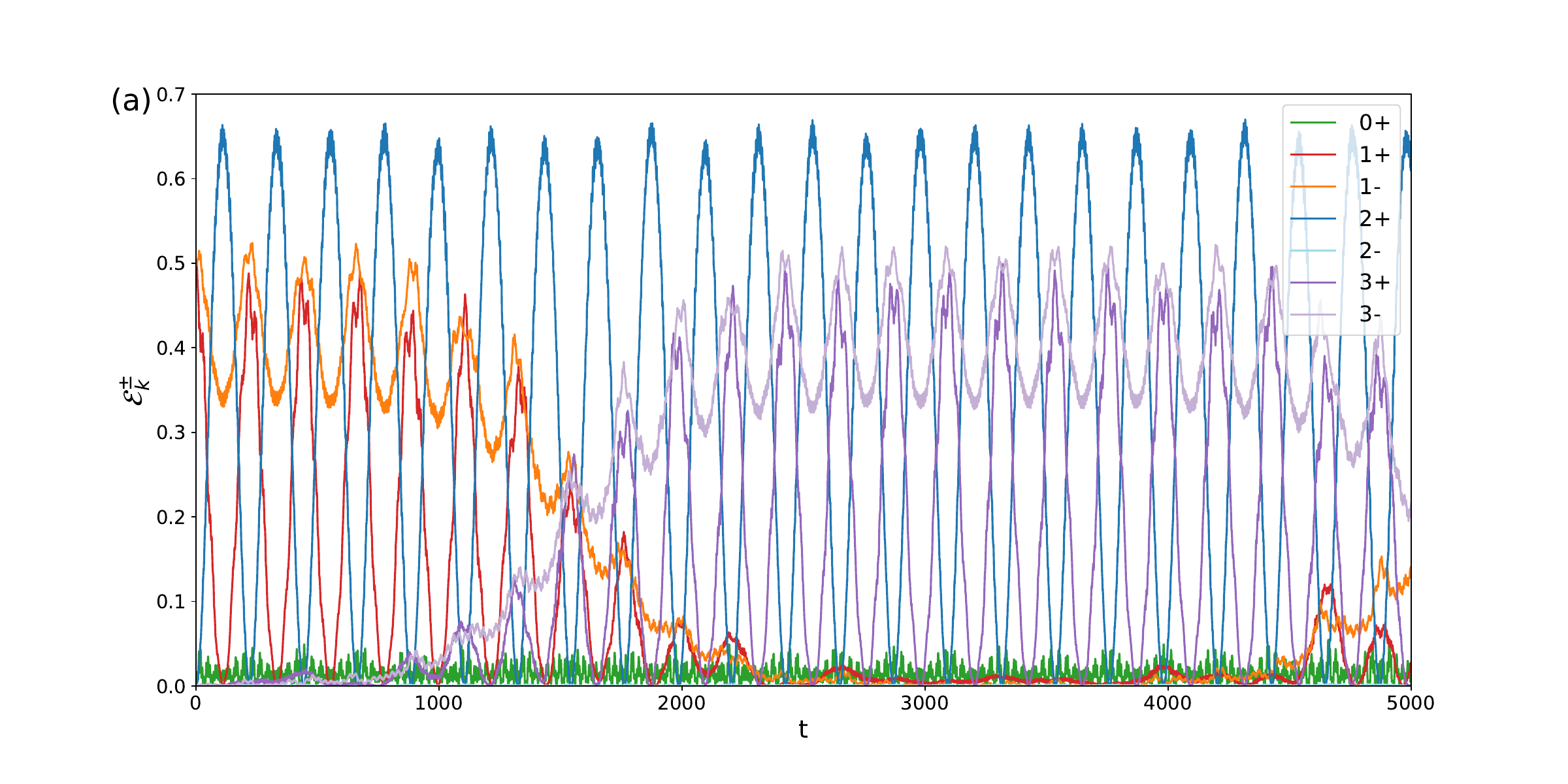} 
	\hspace{0.1cm}
	\includegraphics[scale=0.26,trim={2.9cm 1cm 4cm 1cm},clip]{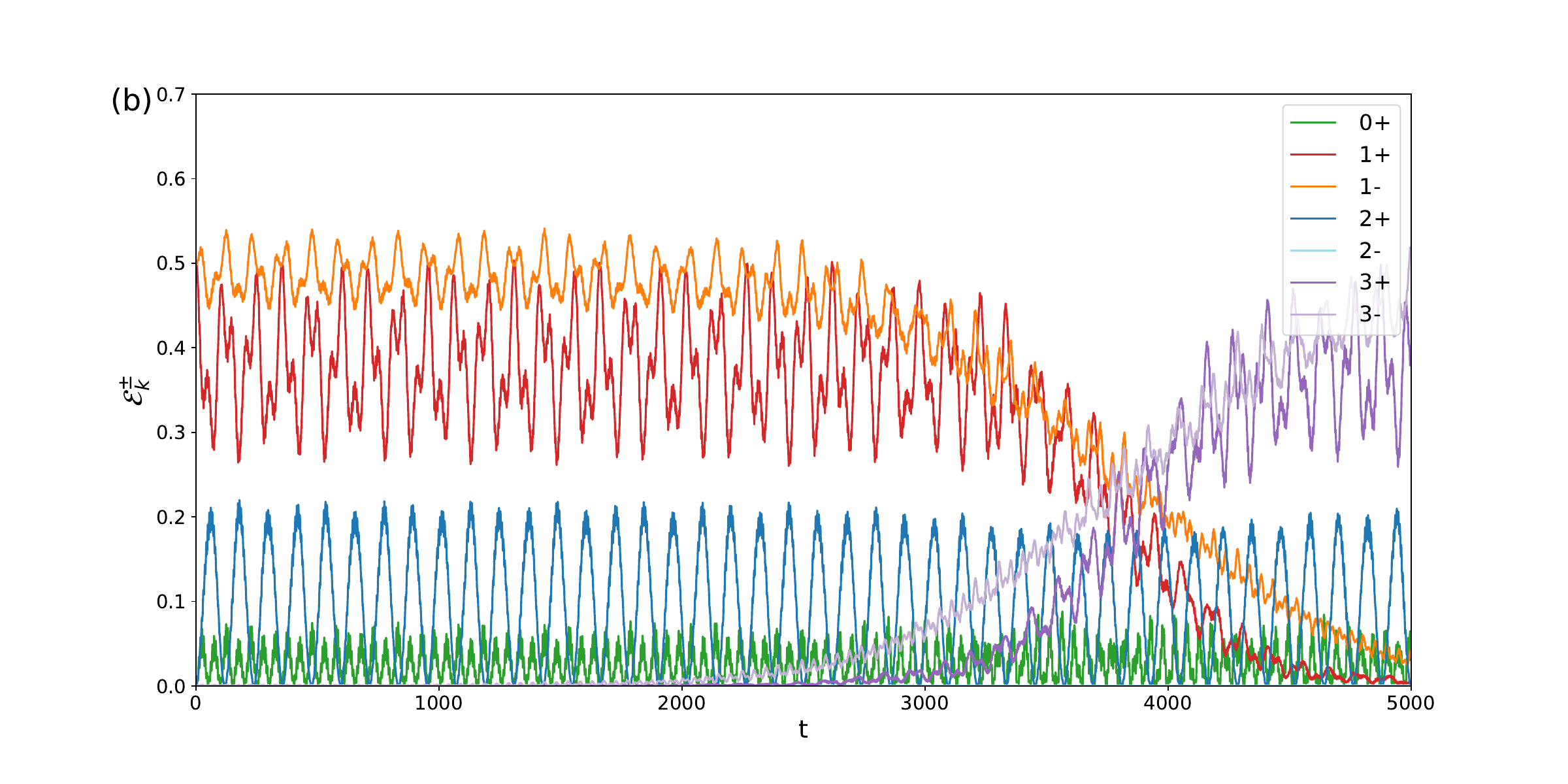}
	\caption{Energy, $\mathcal{E}_k^s = \Omega^s_k \lvert a_k^s \rvert^2$ (with initial condition given in equation \eqref{eq:IC_reso}), as a function of  time for  each normal mode, for the case $N=4$, $\alpha=0.2$ and all other parameters set to one except for $\chi_r$. In (a), $\chi_r$ is selected to solve \eqref{eq:chi}, so that three-wave exact resonances are allowed. In (b), $\chi_r$ has been increased by $15\%$ with respect to the value in a).}
\label{fig:nonlinear_modes}
\end{figure}
In Fig.~\ref{fig:nonlinear_modes}a), we show the evolution in time of the energy,  $\mathcal{E}_k^s = \Omega^s_k \lvert a_k^s \rvert^2$,  associated to each normal modes for the described configuration. As expected from our theory, because the modes $\lbrace 1^-, 1^+ \rbrace$ belong to the resonant triad $(1^-, 1^+, 2^+)$, mode $\lbrace 2^+\rbrace$ starts growing at the expense of the former. Energy is then exchanged periodically in between the modes belonging to the resonant triad. On a longer time scale, it possible to observe that modes $\lbrace 3^-, 3^+ \rbrace$ start growing, and modes $\lbrace 1^-, 1^+ \rbrace$  lose their energy. This mechanism is related to a four-wave resonant interaction, see \eqref{4resonance} and  \eqref{4resonance_s}, which involves the quartet $(1^-, 1^+, 3^-,3^+)$. Interestingly, as the modes $\lbrace 3^-, 3^+ \rbrace$ increase their energy, they start interacting with mode $\lbrace 2^+\rbrace$, thus activating the other resonant triad  $(3^-, 3^+, 2^+)$.  

To further support our analytical results, tests are conducted changing $\chi_r$, so that exact three-wave resonant interactions are detuned, {\it i.e.}, the resonant condition on frequencies is only approximately satisfied. The results are shown in  Fig.~\ref{fig:nonlinear_modes}b), where clearly the exchange of energy between modes $\lbrace 1^-, 1^+ \rbrace$  and mode  $\lbrace 2+ \rbrace$ has been drastically reduced. Interestingly, four-wave resonant interactions are still active, involving  the same quartet $(1^-, 1^+, 3^-,3^+)$ as in the previous case.

\MDB{\section{\label{sec:NKL_FPUT} The recurrence in the nonlinear Kelvin lattice}
As mentioned earlier,  {when $\chi_r$ is set to zero} in equations (\ref{eq_motion}), the system reduces to the standard $\alpha$-FPUT chain \cite{fermi1955studies}. Such a model has been widely studied, and there are excellent reviews on the subject \cite{gallavotti2007fermi,berman2005fermi,carati2005fermi,onorato2023wave,weissert2012genesis}. As is well known, the numerical simulations described in \cite{fermi1955studies} showed some unexpected results: instead of observing the equipartition of energy among the degrees of freedom of the system, the authors observed the quasi-recurrence to the initial state. Therefore, an interesting question to be answered is what happens to the recurrence in the presence of resonators. In this context, although not being the main focus of the present paper,  we have performed  some numerical computations with the aim of assessing the recurrence behaviour in the presence of resonators attached to the chain. As a first step, we have reproduced numerically the original result in \cite{fermi1955studies} by using our model and setting to zero the parameter $\chi_r$ and the initial values of the resonator variables $r_j$ (figure not shown). Next, for the numerical simulations of the {nonlinear} Kelvin lattice, we keep the parameter choices of the original FPUT problem, namely we take $N=32$, $\alpha=0.25$, $\chi_q = \chi_r = 1$ and $m_q = m_r = 1$, consider fixed boundary conditions, and characterize the initial condition by exciting at time $t=0$ only the longest mode of the chain, namely $q_j \propto \sin(\pi j/N)$; in the presence of resonators an initial condition has to be provided also for the variables $r_j$.  Among the infinite number of choices for these initial conditions that can lead to different results, we have picked representative initial conditions using what we believe to be the simplest strategy: to keep exciting only the longest mode (namely $r_j \propto \sin(\pi j/N)$), and to ensure all initial conditions have the same linear part $H_0$ of the initial energy.}

\begin{figure}[htpb!]
	\centering
	\includegraphics[scale=0.3,trim={1cm 1cm 1cm 1cm},clip]{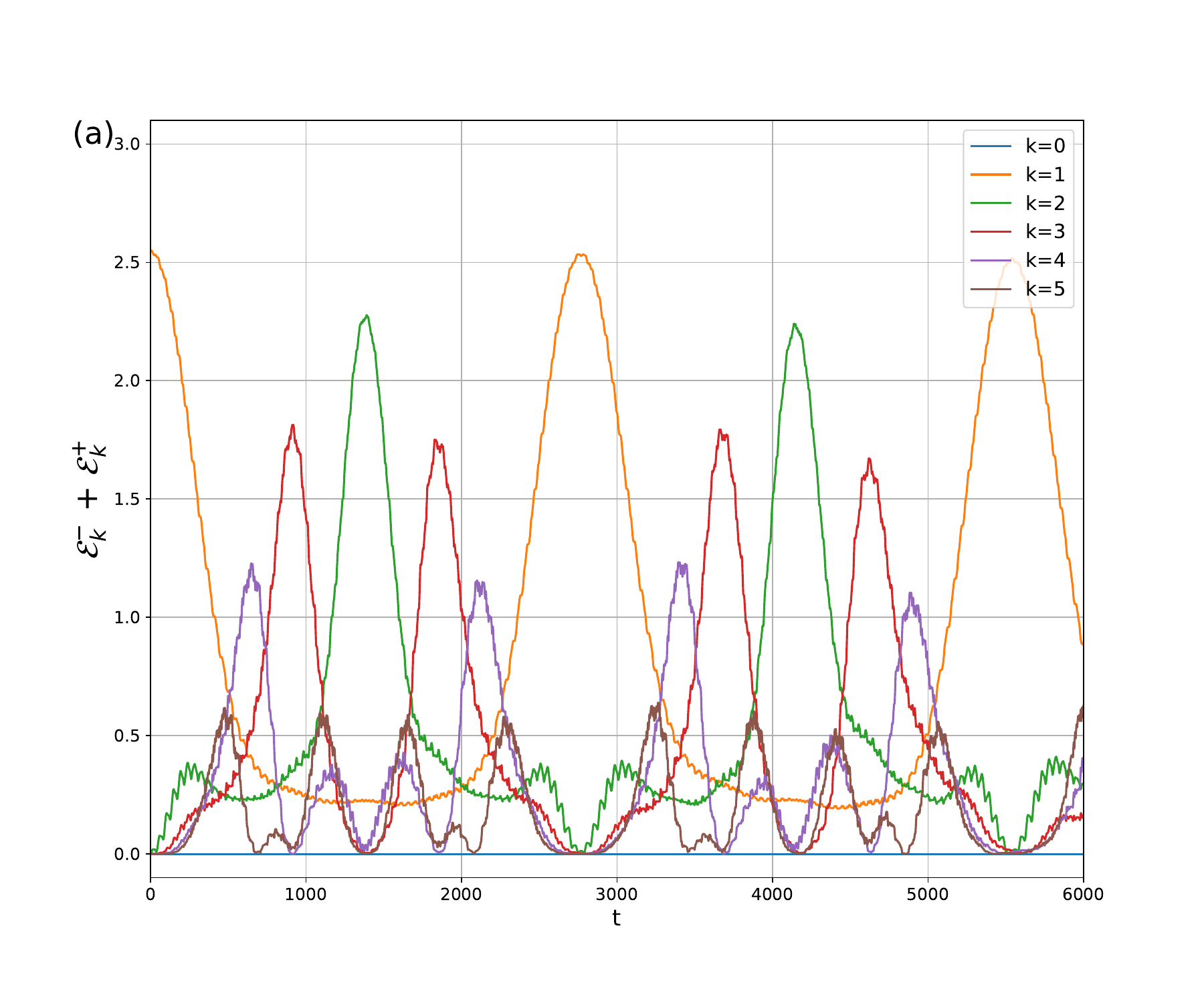} 
	\hspace{0.2cm}
	\includegraphics[scale=0.3,trim={1cm 1cm 1cm 1cm},clip]{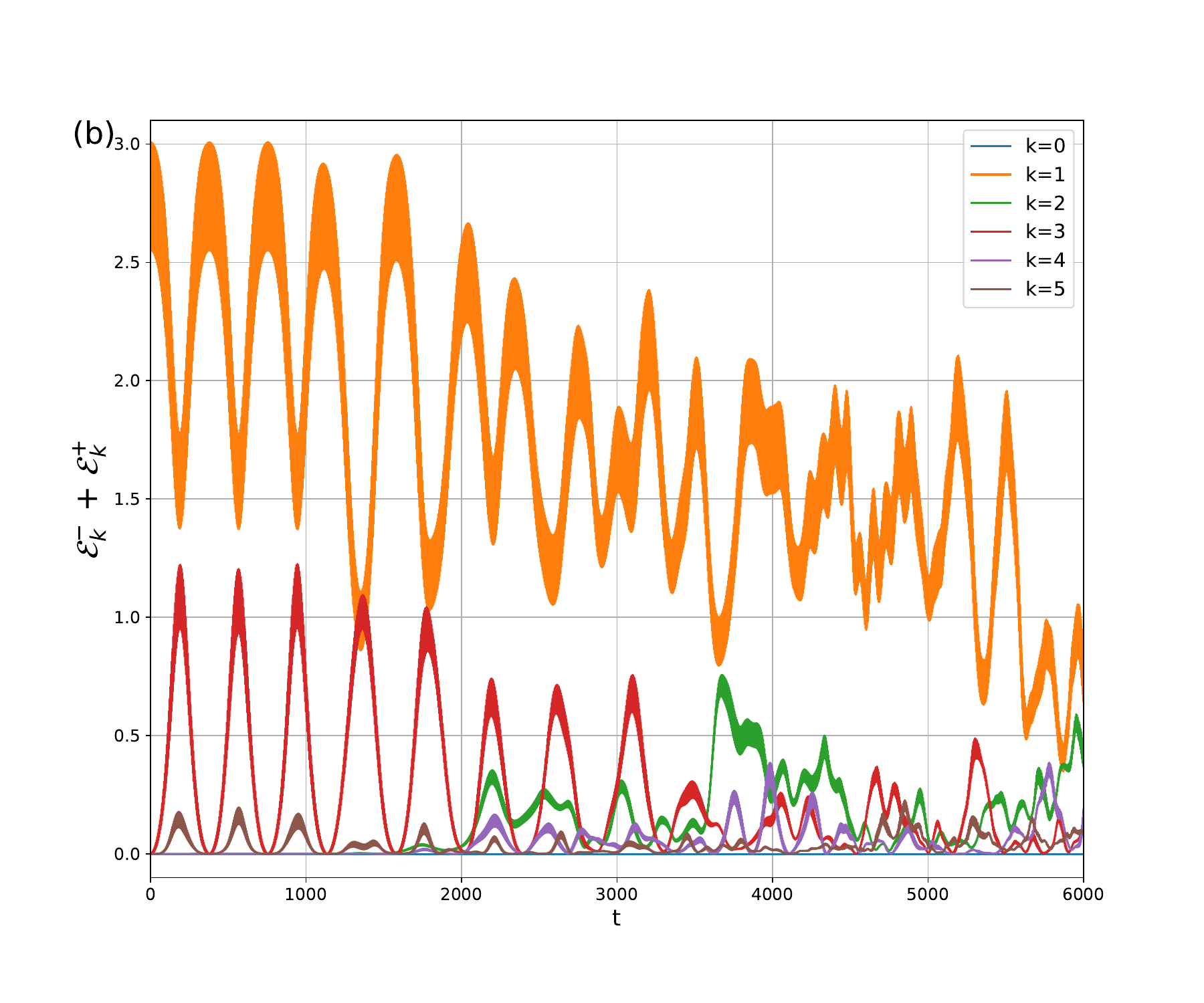}
	\caption{\MDB{Recurrence assessment of the Kelvin lattice for $N=32, \alpha=0.25$ fixed boundary condition case, and initial energy (linear part) $H_0 = 2.5511$. All other parameters are set to one. Evolution of the linear part of the spectral energy, 
 $\mathcal{E}_k^++\mathcal{E}_k^-$, with $\mathcal{E}_k^s = \Omega^s_k \lvert a_k^s\lvert^2$, as a function of  time for the first 5 modes. In (a), the initial condition is characterized by $q_j=\sqrt{N}\sin(\pi j/N) \approx 5.657 \sin(\pi j/N), \quad r_j = \beta_1^- q_j \approx 1.005 \,q_j$, which implies that only spectral mode $a_1^-$ (the longest acoustical mode) is nonzero initially. The system displays recurrence, very much like the classical $\alpha$-FPUT system \cite{fermi1955studies}. In (b), the initial condition is characterized by $q_j=\sqrt{\frac{\mu_1^-}{\mu_1^+}}\frac{\Omega_{1}^-}{\Omega_{1}^+}\sqrt{N}\sin(\pi j/N) \approx 0.2782 \sin(\pi j/N), \quad  r_j = \beta_1^+ q_j \approx -0.9952 \,q_j$, which implies that only spectral mode $a_1^+$ (the longest optical mode) is nonzero initially. The system displays non-recurrent behaviour, as can be seen by the gradual loss of the linear energy to the cubic term of the conserved Hamiltonian (not shown), and the bursting behaviour of the modes' energies.}}
\label{fig:recurrence_study}
\end{figure}

\MDB{Now, to assess the recurrence behaviour, we propose two types of initial conditions based on physical considerations regarding our acoustical/optical splitting of the normal modes. The first type of initial condition is defined so that only the longest acoustical mode is initially excited, namely $r_j = \beta_1^- q_j$, referring to equation \eqref{diffeo}. For this choice, cf.~figure \ref{fig:recurrence_study}(a), the evolution of the system displays a clear recurrent behaviour, strikingly similar to the classical $\alpha$-FPUT system \cite{fermi1955studies}. For our choice of parameters, this gives initially $r_j \approx 1.005 \,q_j$, which makes the cubic term in the Hamiltonian relatively small initially, which could be an explanation for the recurrent behaviour. Analogously, the second type of initial condition is defined so that only the longest optical mode is initially excited, namely $r_j = \beta_1^+ q_j$, again referring to equation \eqref{diffeo}. For this choice, cf.~figure \ref{fig:recurrence_study}(b), the system displays a recurrent behaviour only for a few periods, to then succumb to a gradual loss of linear energy to the cubic term of the conserved Hamiltonian (not shown in the figure), accompanied by bursting behaviour of the other modes' energies. For our choice of parameters, this gives initially $r_j \approx -0.9952 \,q_j$, which makes the cubic term in the Hamiltonian relatively larger initially than in the recurrent case, while still smaller than the linear energy $H_0$, so that during the first stages (before $t=1000$) the linear energy $H_0$ is the main contributor to the Hamiltonian.}

\MDB{Because of the number of parameters involved, the dynamics of our system is very rich, and a detailed study on the recurrence (or non-recurrence) should be performed; however, we find that this interesting research activity is outside the scope of the present paper, and it will be part of future research.}

\section{\label{Conclusions} CONCLUSIONS}
\label{sec:conc}
Starting from the pioneering work of Enrico Fermi and collaborators, \cite{fermi1955studies}, masses and springs have been considered  non trivial toy models suitable for understanding fundamental aspects of propagation of waves in nonlinear dispersive systems. The one dimensional chain with quadratic and cubic nonlinearity has been widely studied in the literature \cite{dauxois2008fermi,berman2005fermi,gallavotti2007fermi} and many results are nowadays available. 
On the other hand, the field of metamaterials is rapidly evolving and it finds application in many different physical problems as for example in electromagnetic, acoustic, seismic,  phononic, elastic, water waves  \cite{hussein2014dynamics,hussein2019advances,miniaci2021hierarchical,schurig2006metamaterial,de2021attenuating}.

In this paper we have tried to understand the propagation properties of a nonlinear chain  coupled with extra resonators that act as a metamaterial. Once the nonlinearity is introduced, the problem becomes much more  complicated from a theoretical point of view; the idea here adopted, and originally developed in the field of Wave Turbulence \cite{nazarenko2011wave}, is to diagonalize the linear system and use the normal variables to describe the nonlinear problem. The resulting model corresponds to two equations, written in Fourier space, which are decoupled at the linear order and the coupling appears in the nonlinear terms. Nonlinearities are expressed as interactions characterized by  rules enforced by the presence of Kronecker deltas, which select the interacting wave numbers via the so-called momentum conditions.  In this framework, exact resonances, i.e. those for which a resonant condition on frequencies is also fulfilled,  play a dominant role in transporting energy between normal modes. 
Because of the quadratic nonlinearity in the equations of motion in this specific case, the interactions are characterized by resonant triads at the leading order in nonlinearity; however, a deeper and accurate analysis reveals that non-resonant triads can generate resonant quartets; this result is well known in the field of surface gravity waves where triads are not resonant \cite{zakharov1968stability}. In addition, we performed numerical simulations of the primitive equations of motion with some specific initial conditions characterized by resonant triads; however, as expected, on a longer time scale, resonant quartets appear naturally in the simulations.
\MDB{Also, in the spirit of the original report by Fermi and collaborators, we  investigated the recurrence behaviour of the nonlinear Kelvin lattice and showed that if only the lowest mode of the acoustical branch is perturbed, recurrence is observed; however, if only the optical branch is perturbed, the recurrence is destroyed. Finally, we calculated (see appendix) the long-wave limit of the nonlinear Kelvin lattice, obtaining a Boussinesq equation that is coupled nonlinearly to a continuum of harmonic oscillators.}

As far as we are aware of, the field of nonlinear metamaterials is only recently being developed. The present approach, innovative in this field, could open up new strategies and new applications in the field.

\begin{acknowledgments}
This publication has emanated from research supported in part by a grant from Science Foundation Ireland under Grant number 18/CRT/6049. For the purpose of Open Access, the authors have applied a CC BY public copyright licence to any Author Accepted Manuscript version arising from this submission. M.O. was funded by Progetti di Ricerca di Interesse Nazionale (PRIN) (Project No. 2020X4T57A), by the European Commission H2020 FET Open Boheme,  grant no. 863179, and by the Simons Foundation, Award 652354 on Wave Turbulence. 
\end{acknowledgments}

\appendix

\MDB{\section{The long wave limit of the nonlinear Kelvin lattice}
\label{sec:appendix}
Zabusky and Kruskal in 1965 \cite{zabusky1965interaction} took the continuum limit of the FPUT chain and obtained the Korteweg-de Vries (KdV) equation. Let us consider a chain with $N$ masses and size $L$. We take the distance between masses to be $a=L/N$. The continuum limit consists of taking the spacing $a$ going to zero and the number of particles to infinity in such a way that $L$ remains constant. Keeping in the expansion for small $a$ the leading order term in dispersion to balance the nonlinearity, the Boussinesq equation is obtained. Assuming waves propagating in only one direction, then the KdV equation is obtained. The procedure is standard and can be found in \cite{ablowitz2011nonlinear}. 
We now start from equation (\ref{eq_motion}) and perform the same procedure as just explained, so that $q_j(t)$ and $r_j(t)$ become continuous functions of $x$ and $t$ in the following way:
\begin{equation}
\begin{split}
&q_j(t)\rightarrow q(x,t)\,,\\
&q_{j \pm 1}\rightarrow q(x\pm a,t)\,,\\
&r_j(t)\rightarrow r(x,t)\,.
\end{split}
\end{equation}
We then Taylor expand $q(x\pm a,t)$ up to fourth-order for small $a$ and plug the result in equations   (\ref{eq_motion}) to obtain:
\begin{subequations}
	\label{eq_motion_cont_NL}
	\begin{align}
		q_{tt}-c^2q_{xx}&=b^2q_{4x}+\gamma q_xq_{xx}
		-\frac{\chi_r}{m_q}(q-r)-\frac{\alpha}{m_q}(q-r)^2\,,
		\label{eq_motion_cont_q_NL}\\
		r_{tt}&=\frac{\chi_r}{m_r}(q-r)+\frac{\alpha}{m_r}(q-r)^2\,,
		\label{eq_motion_cont_r_NL}
	\end{align}
\end{subequations} 
where
\begin{equation}
	\label{cbg}
	c^2=\frac{\chi_q}{m_q}a^2\,, \qquad
	b^2=\frac{1}{12}\frac{\chi_q}{m_q} a^4\,, \qquad 
	\gamma=\frac{2\alpha}{m_q}a^3\,.
\end{equation}
Note that, setting to zero the field $r$ (which represents the resonators' displacement variable in the continuum limit), takes us back to the standard Boussinesq equation, from which the KdV equation can be derived. The coupled system in (\ref{eq_motion_cont_NL}) constitutes a new model and deserves a detailed study which, however, is outside the scope of the present paper.
}

\bibliography{sample.bib}

\end{document}